\documentclass[letterpaper,twocolumn,10pt]{article}
\usepackage{usenix,epsfig,endnotes}
\usepackage{subfigure}
\usepackage{url}
\usepackage{hyperref}

\begin{document}

\date{}

\title{\Large \bf TaNG: Modeling Packet Classification with TSS-assisted Neural Networks on GPUs}

\author{
{\rm Zhengyu~Liao, Shiyou~Qian$^*$,}\\ 
{\rm Jian~Cao, Guangtao~Xue}\\
Shanghai Jiao Tong University
\and
{\rm Zhonglong~Zheng, Minglu~Li}\\
Zhejiang Normal University
\and
{\rm Jiange Zhang}\\
Huawei Technologies
}

\maketitle

\thispagestyle{empty}

\subsection*{Abstract}
Packet classification is a core function in software-defined networks, and learning-based methods have recently shown significant throughput gains on large-scale rulesets.
However, existing learning-based approaches struggle with overlapping rules, leading to incomplete model coverage or excessive rule replication. Their limited GPU integration also hampers performance with large-scale rulesets.
To address these issues, we propose TaNG, which utilizes a single neural network trained on multi-dimensional features to ensure complete coverage without duplicating rules.
TaNG employs a semi-structured design that combines the neural network model with a tuple space, reducing model complexity. Furthermore, we develop a mechanism based on the semi-structure for rule updates. Finally, we implement a CPU–GPU hybrid streaming framework tailored for learning-based methods, further enhancing throughput.
On our GPU-based classification framework with 512k rulesets, TaNG achieves 12.19x and 9.37x higher throughput and 98.84x and 156.98x higher performance stability compared to two state-of-the-art learning methods NuevoMatch and NeuTree, respectively.

\section{Introduction}

Packet classification is a core function in modern network systems, supporting applications such as security enforcement, network monitoring, and multimedia communications \cite{survey,7164675,832499}. It assigns incoming packets to actions based on multi-field rulesets, enabling fine-grained traffic control.
As network bandwidth \cite{8535476, CHEN2020312} and traffic volumes grow \cite{1402467}, packet classifiers should sustain high throughput on rulesets at a million-scale, which is challenging due to large ruleset sizes and diverse traffic patterns.
The emergence of software-defined networking (SDN) \cite{hakiri2014software, 7289347,8360763,DEB2022108802} further increases the difficulty, as classifiers must adapt dynamically to changing policies without sacrificing performance. Addressing these requirements calls for packet classification approaches that balance speed, scalability, and update efficiency, which remains a critical open problem in high-bandwidth networking.


Packet classification algorithms can be broadly divided into TCAM-based hardware solutions \cite{10.5555/1875239,10.1145/2079296.2079323,6165260,7110119,7005545,7707346,7061970,8008845} and software-based approaches \cite{bitcuts,daly2018bytecuts,8901884,CutSplit2018,TSS1999,TupleMerge2019,li2020tuple,9982296,pt-tree,dbtable2024}. Software methods are generally more flexible and cost-effective for large, dynamic rulesets and fall into decision-tree \cite{gupta1999packet,singh2003packet,qi2009packet,vamanan2010efficuts,fong2012parasplit,li2013hybridcuts,he2014meta,2016sorted,bitcuts,daly2018bytecuts,8901884,CutSplit2018} and hash-based \cite{TSS1999,pfaff2015design,TupleMerge2019,ZHANG2022108630} classes. Decision trees recursively partition the search space, while hash-based methods use rule prefixes for direct lookups. Both suffer drawbacks: decision trees face rule replication, and hash methods incur extensive table accesses, limiting scalability. Hybrid approaches \cite{li2020tuple,9982296,pt-tree} partially mitigate these issues but cannot fully overcome them. Some works port software-based approaches to GPUs \cite{10.1145/2674005.2674990,AbbasiS20,9078110,10431734}, yet their comparison- or bitwise-heavy operations yield poor parallel efficiency.

The rapid advancement of machine learning has introduced a new paradigm for packet classification. Some studies have focused on using techniques like reinforcement learning \cite{9904958} to optimize existing data structures \cite{liang2019neural,liu2022hybridtss}. A more disruptive approach, however, involves replacing traditional data structures entirely with neural network models, known as learned indexes \cite{NeuvoMatch2020,LI2024110745,neutree2025}. Most learned index methods for packet classification are based on the Recursive Model Index (RMI) framework \cite{rmi2018}, which constructs hierarchical neural network models along different rule dimensions to create a mapping from packets to rules. As these models primarily involve a fixed number of floating-point computations, learned indexes can achieve significantly higher and more stable performance compared to traditional software methods, while also being well-suited for GPU deployment.


Existing RMI-based packet classification approaches face challenges in both model design and hardware integration. On the modeling side, RMIs struggle with high-dimensional and overlapping rule spaces, prompting methods like NuevoMatch \cite{NeuvoMatch2020} to isolate some overlapping rules into a remainder set and NeuTree \cite{neutree2025} to allow rule replication for non-overlapping buckets. However, these strategies introduce overhead, reducing model coverage or increasing search costs, especially for skewed rulesets. On the hardware side, RMI hierarchies create branch-selection bottlenecks, and current methods underutilize GPUs. 
These limitations underscore the need for an approach that efficiently handles large, overlapping rulesets while fully leveraging GPU floating-point computation capabilities.

In this paper, we propose TaNG, a novel learning-based packet classification approach. 
TaNG does not rely on the RMI framework, instead, it reformulates the problem from the RMI-based regression paradigm into a classification paradigm. By utilizing the multi-dimensional features of rules, a single, unified neural network classification model is constructed. This model achieves complete ruleset coverage without replicating any rules and is ideally suited for GPU-accelerated inference.
To reduce model complexity, we introduce a semi-structured design by incorporating the tuple space as a middleware between the model and rules. Building on this design, we develop a mechanism for rule updates, avoiding frequent model retraining. Moreover, we implement a CPU-GPU streaming classification framework tailored for learning-based methods, leveraging efficient floating-point computation and parallelism to achieve 
high throughput.


We conduct a comprehensive evaluation of TaNG, assessing its classification performance, maintenance overhead, and rule update capability. The experiments use rule and packet sets generated by classbench-ng \cite{classbench-ng}, as well as real-world traffic, with rulesets scaling up to 512K entries. In comparative experiments, we primarily evaluate TaNG against two representative learning-based methods, NuevoMatch \cite{NeuvoMatch2020} and NeuTree \cite{neutree2025}, implemented under our CPU-GPU streaming classification framework. The results demonstrate that TaNG achieves substantial improvements in classification throughput: for the 512K ruleset, its average throughput is 12.19x and 9.37x higher than NuevoMatch and NeuTree, respectively, while its throughput stability across different rulesets is 98.84x and 156.98x higher. We further compare TaNG with three representative traditional methods—PSTSS \cite{pfaff2015design}, CutSplit \cite{CutSplit2018}, and DBTable \cite{dbtable2024}. We observe that its average throughput is 28.92x, 13.45x, and 3.76x higher, respectively.

The main contributions of this work are as follows:
\begin{itemize}
    \item We propose a single unified neural network classification model that leverages the multi-dimensional features of rules, achieving complete ruleset coverage without rule replication.

    \item We introduce a semi-structured design that leverages tuple spaces as a middleware, reducing model complexity, and implement a post-verification mechanism to improve classification accuracy.

    \item We implement a update mechanism based on the semi-structured design and a CPU-GPU hybrid framework to accelerate floating-point inference tailored for learning-based methods.
\end{itemize}

The remainder of this paper is organized as follows. Section \ref{sec:back_relat} reviews background and related work. Section \ref{sec:moti} presents the motivation for this work. Section \ref{sec:design} describes the design of TaNG, followed by Section \ref{sec:op}, which details the classification workflow and update mechanism. Section \ref{sec:imple} covers the neural network implementation, and Section \ref{sec:eval} presents the experimental evaluation. Section \ref{sec:limi} discusses limitations and future directions, and Section \ref{sec:conclu} concludes the paper.

\section{Background and Related Work}
\label{sec:back_relat}

In this section, we present a formal description of the packet classification problem and review existing packet classification approaches.

\subsection{Packet Classification Problem}

Network packet classification aims to enable differentiated packet processing using a classifier or ruleset, which consists of predefined rules. Each rule, denoted as $R = \{r_1, r_2, \dots, r_F\}$, comprises $F$ components, where $r_i$ specifies a condition on field $f_i$. Each rule is associated with a priority and an action. Common fields include source IP (SIP), destination IP (DIP), source port (SP), destination port (DP), and protocol (PRO). Conditions are typically expressed as prefixes for IP addresses, exact or wildcard ($*$) matches for protocols, and ranges for ports. Each packet, denoted as $P = \{p_1, p_2, \dots, p_F\}$, also has $F$ fields. Each field has $d$ bits, with values in $[0, 2^d - 1]$. Packet classification typically requires identifying the highest-priority rule that matches a given packet $P$ and executing its associated action. 

\subsection{Hardware-based Methods}

Hardware-based solutions for packet classification are dominated by Ternary Content-Addressable Memories (TCAMs), which achieve line-rate performance~\cite{10.5555/1875239,10.1145/2079296.2079323,6165260,7110119,7005545,7707346}.
TCAMs compare an input key against all entries in parallel, supporting exact, prefix, and range matches via three entry states (0, 1, “don’t care”).
This makes TCAMs particularly suitable for packet classification. However, despite their line-rate performance, TCAMs are costly, and they impose high complexity on rule updates, which limits scalability. To address this, recent work has explored algorithmic and hardware-based alternatives~\cite{7061970,8008845}. While TCAMs’ intrinsic limitations make them less suitable for SDN deployments, they have nonetheless inspired a variety of software-based solutions.


\subsection{Traditional Software-based Methods}

Traditional software-based packet classification methods primarily rely on either decision tree-based or tuple-based schemes, each with distinct tradeoffs between lookup speed and update efficiency. Decision tree-based methods recursively partition the rule space into subregions until each region contains few rules, enabling efficient lookups by traversing a single leaf. However, rule replication across partitions hinders updates. HiCuts~\cite{gupta1999packet} introduced uniform space partitioning, later extended by HyperCuts~\cite{singh2003packet} with flexible cuts and HyperSplit~\cite{qi2009packet} with replication-aware splitting. EffiCuts~\cite{vamanan2010efficuts} builds multiple independent trees to further reduce replication, while PartitionSort~\cite{2016sorted} eliminates replication via multi-key binary search trees. Fine-grained techniques like BitCuts~\cite{bitcuts} and ByteCuts~\cite{daly2018bytecuts} optimize memory usage and replication, and CutSplit~\cite{CutSplit2018} uses a hybrid cut-and-split design to reduce rule replication and tree depth.


Tuple-based methods, in contrast, divide rules into tuples based on prefix lengths and index them via hash tables, inherently supporting efficient updates. The original TSS~\cite{TSS1999} suffers from exhaustive tuple probing, which PSTSS~\cite{pfaff2015design} mitigates using priority-based ordering. TupleMerge~\cite{TupleMerge2019} and DynamicTuple~\cite{ZHANG2022108630} merge similar tuples to reduce lookup overhead, though collisions remain a risk. DBTable~\cite{dbtable2024} further maps all rules into a single hash table for higher lookup efficiency. Hybrid schemes like CutTSS~\cite{li2020tuple}, TupleTree~\cite{9982296}, and PT-Tree~\cite{pt-tree} combine tree and tuple ideas to balance lookup performance and update efficiency, though some require multiple leaf accesses.

Besides software optimizations, several studies leverage GPU parallelism to accelerate traditional software-based methods~\cite{10.1145/2674005.2674990,AbbasiS20,9078110,10431734}. These approaches parallelize decision trees, hash tables, or hybrid algorithms across GPU threads to increase throughput. However, sequential comparisons and bitwise operations in these algorithms are not well suited for GPU execution. 
Overall, most decision tree methods achieve fast lookups but struggle with updates due to replication, whereas tuple-based methods usually enable efficient updates but may exhibit unstable lookup performance.



\subsection{Learning-based Methods}



Learned indexing \cite{rmi2018} represents a paradigm shift in data management, integrating machine learning techniques into the design of indexing structures. Existing research on packet classification in this area has largely followed two directions: one leverages machine learning to tune and optimize parameters of existing index structures for improved efficiency \cite{liang2019neural,liu2022hybridtss}, while the other replaces traditional index structures with predictive models that directly estimate data positions \cite{NeuvoMatch2020,LI2024110745,neutree2025}. The latter model-based prediction approach has attracted particular attention for two key reasons. First, by compressing conventional data structures into compact neural network models, it reduces memory consumption and enables more stable lookup performance across diverse workloads. Second, it transforms CPU-intensive logical operations into floating-point computations, enabling learning-based methods more suitable for GPU acceleration on large-scale rulesets.

Optimization-oriented approaches typically leverage reinforcement learning \cite{9904958}, such as NeuroCuts \cite{liang2019neural}, which optimizes decision tree construction, and HybridTSS \cite{liu2022hybridtss}, which improves the structure of TSS. In contrast, replacement-oriented approaches are primarily built upon the RMI framework \cite{rmi2018}. RMI treats indexing as a regression problem: given a key, the model predicts its position in the sorted dataset. Conceptually, this corresponds to regressing the polyline formed by key–position pairs. RMI builds a hierarchy of models, where each intermediate model directs a submodel in the next layer, and the final-layer model outputs a predicted position. Since the dataset are ordered, correctness is guaranteed by performing a local search around the predicted index to recover the exact position.

NuevoMatch\cite{NeuvoMatch2020} and NeuTree \cite{neutree2025} are representative learning-based approaches built upon RMI. NuevoMatch introduces the Range Query Recursive Model Index (RQ-RMI) model, which enables the mapping of packets to non-overlapping rules along a single field. This approach was the first to translate a ruleset into neural network models, reducing rule storage overhead and improving throughput on large-scale rulesets. NeuTree further addresses the limitation of NuevoMatch in handling overlapping rules by introducing a bucket division mechanism. By allowing rule replication across buckets to achieve isolation between buckets, NeuTree enables complete model coverage and achieves further throughput improvements.

\begin{figure}[tbp]
  \centering
  \begin{minipage}[t]{0.49\linewidth}
    \centering
    \includegraphics[width=\linewidth]{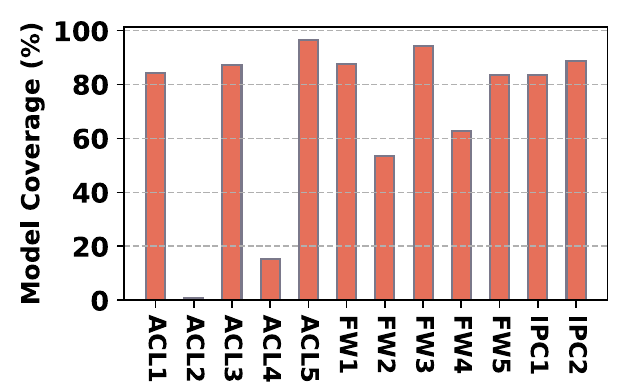}
    \caption{The model coverage of NuevoMatch on 12 512k rulesets.}
    \label{fig:nm}
  \end{minipage}%
  \hfill
  \begin{minipage}[t]{0.49\linewidth}
    \centering
    \includegraphics[width=\linewidth]{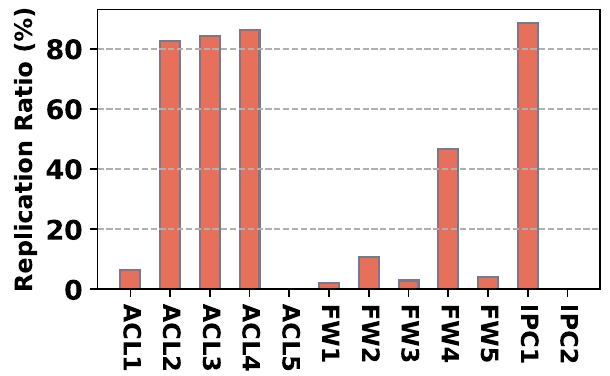}
    \caption{The replication ratio of NeuTree on 12 512k rulesets.}
    \label{fig:nt}
  \end{minipage}
\end{figure}

\section{Motivation}
\label{sec:moti}

Despite the promising results of learning-based methods on large-scale rulesets, they still suffer from several limitations, primarily in terms of model framework and hardware–software integration.
Specifically, on the model framework side, RMI-based methods construct regression models along a single field for a set of rules, requiring that the rules in the set have non-overlapping constrained intervals on that field. If rules overlap, packets in the overlapping regions correspond to multiple rules, resulting in a one-to-many mapping that the RMI framework cannot directly handle. 
Given a field with a fixed value domain, rule overlapping is inevitable and becomes more pronounced with an increase in the number of rules, posing a significant challenge for RMI-based methods.

To address this issue, NuevoMatch partitions the ruleset into multiple non-overlapping subsets and constructs an RQ-RMI model for each subset. While this approach enables the rule-to-model conversion, the construction of a large number of models can significantly degrade overall performance. 
To mitigate this overhead, NuevoMatch limits the number of models and constructs a reminder set, delegating the rules within it to a traditional approach for processing.
However, this strategy leads to a substantial drop in model coverage when facing skewed rulesets. Figure \ref{fig:nm} illustrates the model coverage of NuevoMatch across various 512K rulesets, using the setup in Section \ref{sec:ex_set}. Notably, for highly overlapping rulesets, coverage can fall below 60\%, and on ACL2 it drops to as low as 0.81\%, effectively causing NuevoMatch to degenerate into the traditional approach.

In contrast, NeuTree partitions the value domain of a given field into non-overlapping intervals, each corresponding to a bucket, allowing the model to directly predict the bucket index. 
Each bucket holds all rules whose constraint interval on the given field overlap with the bucket’s responsible interval, ensuring correct rule lookup within the bucket. 
While this design achieves full model coverage, it introduces rule replication, which can be severe when many rules specify large constraint intervals. To mitigate this, NeuTree further divides rules into four subsets based on interval size, reducing but not completely eliminating replication.
Extensive rule replication not only increases maintenance overhead but also degrades the efficiency of bucket-level rule lookup, undermining the stability of classification performance. Figure \ref{fig:nt} shows the rule replication rate of NeuTree across 12 512K rulesets. On ACL2, ACL3, ACL4, and IPC2, the replication rate exceeds 80\%, highlighting a significant limitation of this approach.

On the other hand, existing learning-based packet classification methods have yet to fully exploit modern GPUs for floating-point computation and large-scale parallelism, limiting the performance. Specifically, NuevoMatch leverages the AVX-512 instruction set on CPUs, yet its floating-point throughput remains substantially lower than GPUs. NeuTree translates floating-point operations into integer and shift computations for parallel execution on FPGAs, but this sacrifices prediction accuracy, leading to additional search overhead.
Moreover, both inherit a fundamental bottleneck from the RMI architecture: its recursive hierarchy may require traversing different branches at each layer, increasing parallelization costs. 
Taken together, these limitations highlight the need for an efficient software-hardware co-design for learning-based methods to handle massive rulesets and ultra-high throughput.


\section{TaNG Design}
\label{sec:design}

This section introduces TaNG, focusing on its paradigm shift, semi-structured design, and overall workflow.


\subsection{Basic Idea}



The core idea of TaNG is to construct a single neural network classification model as a predictor. 
To address the limitations of existing learning-based approaches, we move beyond the paradigm of building RMI regression models on individual dimensions and instead construct a single classification model over the entire ruleset in high-dimensional space. Essentially, packet classification is a point classification problem in high-dimensional space.
Thus, this design inherently resolves model coverage issues and simultaneously prevents rule replication.
In addition, adopting a non-recursive single-model structure enhances GPU efficiency by fully leveraging the floating-point computation capabilities of modern GPUs.

However, realizing this idea faces three key challenges. First, how to effectively model the mapping from packets to the matching rules within the entire ruleset? Second, how to control model complexity to maintain high throughput? Third, how to support rule updates without compromising efficiency? To this end, we first reformulate the problem from an RMI-style regression paradigm to a classification paradigm, enabling the construction of a single model. We then introduce a TSS-assisted neural network design, referred to as a semi-structured model, which reduces model complexity while supporting rule updates.



\begin{table}[tbp]
\renewcommand\arraystretch{1}
\centering
  \caption{Rules with two fields, each occupying 3 bits.}
  \label{tab:rules}
  \begin{tabular}{c|c c c | c } \hline
  \textbf{Rule ID} & \textbf{Priority} & \textbf{Field $X$} & \textbf{Field $Y$} & \textbf{Tuple} \\ \hline
 $R_1$ & 1 & 000 & 011 & $T_1$ (3, 3) \\
 $R_2$ & 2 & 000 & 101 & $T_1$ (3, 3) \\
 $R_3$ & 3 & 00* & 11* & $T_2$ (2, 2) \\
 $R_4$ & 4 & 110 & *** & $T_3$ (3, 0) \\
 $R_5$ & 5 & 111 & *** & $T_3$ (3, 0) \\
 $R_6$ & 6 & *** & 011 & $T_4$ (0, 3) \\
 $R_7$ & 7 & *** & 010 & $T_4$ (0, 3) \\ 
 $R_8$ & 8 & 0** & 0** & $T_5$ (1, 1) \\ \hline
  \end{tabular}
\end{table}

\subsection{Problem Reformulation}

In packet classification, each rule is defined by range constraints across multiple dimensions, effectively forming a hyper-rectangle in a high-dimensional space. A significant complication is that these hyper-rectangles can overlap, which makes building a classification model based solely on rule constraints highly challenging. Fortunately, rules are also associated with priority information. This priority scheme ensures that any point (corresponding to a packet) in the high-dimensional space is unambiguously matched by a rule. In other words, every point has a unique class label. 
Naturally, we can reformulate the packet classification problem as a multi-class classification task in a high-dimensional space, allowing us to construct a single multi-class model over the entire ruleset to map packets directly to rules.

\begin{figure}[tbp]
  \centering
  \begin{minipage}[t]{0.49\linewidth}
    \centering
    \includegraphics[width=\linewidth]{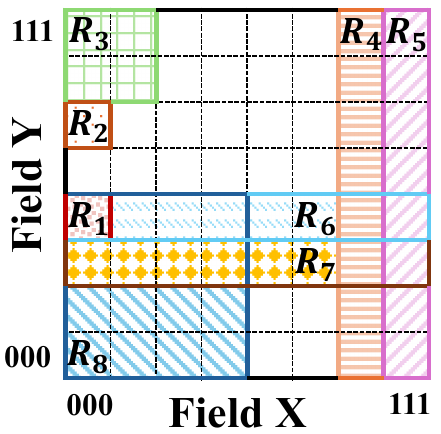}
    \caption{The effective constraint space of rules.}
    \label{fig:rules}
  \end{minipage}%
  \hfill
  \begin{minipage}[t]{0.49\linewidth}
    \centering
    \includegraphics[width=\linewidth]{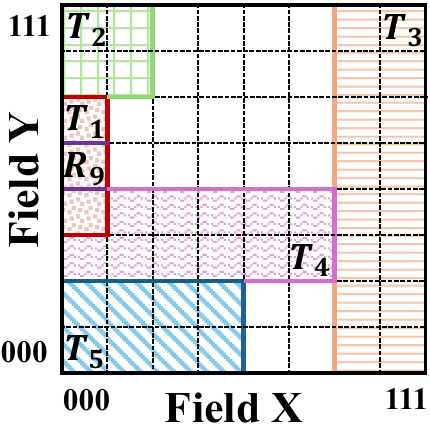}
    \caption{The example tuple boundaries learned by a model.}
    \label{fig:tuples}
  \end{minipage}
\end{figure}

Figure \ref{fig:rules} provides a schematic illustration of this principle in a two-dimensional space. The figure depicts 8 rules ($R_1-R_8$) in Table \ref{tab:rules}, with the effective decision space of each rule identified by different color patterns. It is evident that although the defined spaces for the rules overlap, the priority mechanism carves out clear, non-overlapping decision boundaries. For example, while the space for rule $R_1$ overlaps with that of $R_6$ and $R_8$, $R_1$'s higher priority dictates that the contested area is assigned exclusively to it. Essentially, every rule possesses a distinct and unambiguous boundary in the high-dimensional space. 
This implies that each rule can be regarded as a class, and it is feasible to train a multi-class model that learns the actual boundaries of rules to perform packet classification.

However, in packet classification, the number of rules can reach millions.
Treating each rule as an individual class and constructing a fully unstructured, pure neural network model would make the model prohibitively complex, hindering high-throughput performance. Moreover, training such a model becomes increasingly difficult, and prediction accuracy is hard to guarantee. In addition, any update to the ruleset may alter class boundaries or class numbers, forcing model reconstruction and making efficient updates infeasible. 
To address this, we aim to leverage the structured design from traditional methods in combination with neural network models, using conventional data structures as a middleware to decouple the model from rules and help solve the aforementioned issues.
We refer to this as a semi-structured model.



\subsection{Semi-structured Model}
\label{sec:semi-structure}

To make our single-model approach practical, the structured middleware must decouple the predictor from rules, preserve exact semantics, bound lookup cost, and remain easy to build and update. Specifically:
\begin{enumerate}
\renewcommand{\labelenumi}{(\arabic{enumi})}
    \item \textbf{Class compression with fidelity}: drastically reduce prediction targets while preserving the packet-to-rule mapping via a surjective rule-to-middleware mapping, ensuring the highest-priority match is accessible.
    \item \textbf{Low-cost post-inference search}: ensure bounded, cache-friendly search within the middleware so that throughput is dominated by inference, while guaranteeing correctness under probabilistic predictions. 
    \item \textbf{Simple construction and fast updates}:
    ensure the middleware is easy to build and update, supporting immediate inserts/deletes without altering its structure, thereby avoiding frequent model retraining.
\end{enumerate}

TSS \cite{TSS1999} serves as an ideal middleware that meets these requirements. First, it enables class compression while preserving one-to-one mapping by grouping rules based on prefix lengths, with each group forming a tuple uniquely identified across multiple field prefix lengths. Each rule maps to exactly one tuple, realizing a surjective mapping. From a spatial perspective, each tuple merges the constraint spaces of its contained rules, converting packet-to-rule mapping into packet-to-tuple mapping and reducing the number of categories, thus lowering model complexity. For example, in typical 5-tuple rulesets, tuples built from SIP and DIP fields are usually fewer than 400 under 256k and 512k rulesets (Section \ref{sec:ex_infor}). Figure~\ref{fig:tuples} shows example tuple boundaries from Table~\ref{tab:rules}, with $T_4$ corresponding to the union of $R_6$ and $R_7$.


Second, searching within a tuple is fast. Rules in a tuple are truncated according to the tuple’s prefix-length signature, and then organized into a hash table indexed by these truncated values, enabling constant-time lookups.
For example, consider a 3-bit field $f$ in a tuple $T$ with a prefix length of 2. If a rule with $r_f = 11*$, the truncated value of $r_f$ in $T$ is $110$.
This ensures that post-inference search remains lightweight, so classification throughput is bounded by model inference rather than search overhead. Furthermore, when model predictions are incorrect, sequential searches across other tuples can guarantee correctness, as detailed in Section \ref{sec:classification}, which is significantly more efficient than directly searching the entire ruleset.

Finally, TSS is simple to construct and supports immediate rule updates without retraining the model. Since tuple construction depends only on rule prefix lengths, it requires neither complex computation nor global coordination, as validated in Section~\ref{sec:ex_constru}. Moreover, updates are efficient: a tuple is located via prefix information, and the rule is updated directly in its hash table. Because the model is decoupled from rules, updates within a tuple have limited impact, thus preserving stable performance. For example, inserting $R_9=\{X=000,Y=100\}$ with prefix $(3,3)$ maps directly to $T_1$. When the model-learned boundary of $T_1$ is as shown in Figure~\ref{fig:tuples}, the model continues to function correctly.


However, conventional TSS update mechanisms may create or remove tuples, which invalidate the model. To address this, we introduce a restricted tuple update strategy that guarantees immediate rule updates without model retraining (Section~\ref{sec:update}). Building on these properties, we adopt TSS as the middleware and introduce a semi-structured model that decouples the classifier from rules to assist classification. Thus, by integrating the traditional data structure, the model predicts tuples rather than individual rules, thereby reducing complexity while enabling efficient updates.


\subsection{Workflow Overview}
\begin{figure}
    \centering
    \includegraphics[width=1\linewidth]{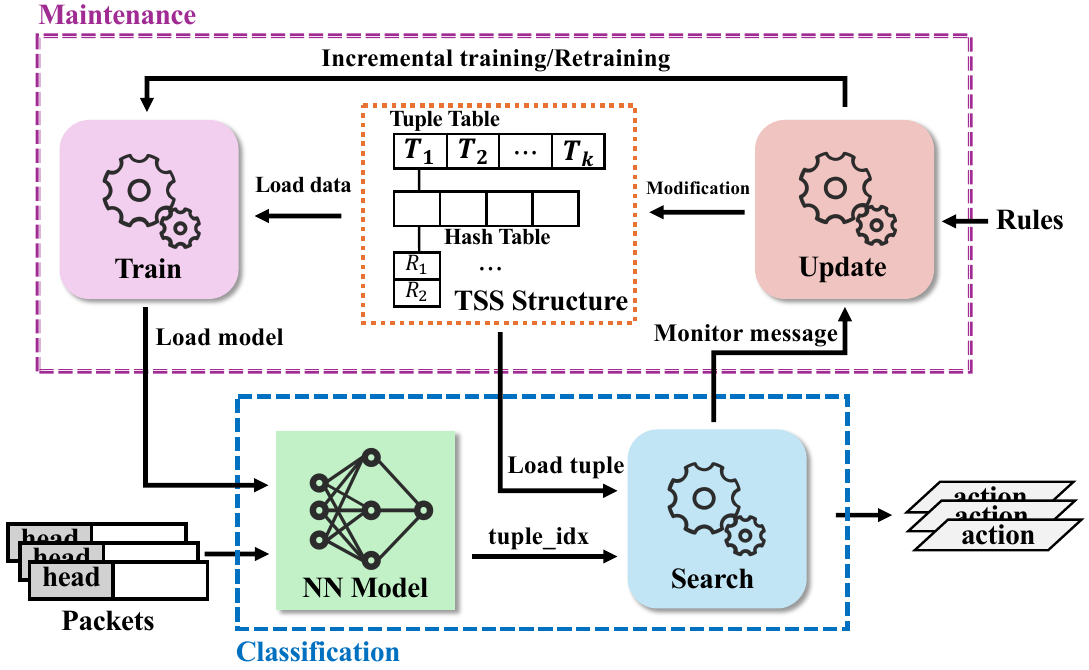}
    \caption{TaNG classification and maintenance workflow.}
    \label{fig:workflow}
\end{figure}

Building on the above semi-structured design, we present an overview of the TaNG classification and maintenance workflow, as shown in Figure~\ref{fig:workflow}. 
The classification workflow executes packet lookups in real time. Given a packet header, the neural network model predicts the target tuple, which is then passed to the search engine. The search engine performs a constant-time hash lookup within the tuple and returns the action of the highest-priority matching rule. 
Section \ref{sec:classification} provides a detailed walk-through of this classification workflow.

The maintenance workflow manages both the construction and evolution of the classifier. It begins by creating the TSS structure for a given ruleset, from which training data is generated and fed into the training engine. The resulting neural network model is then deployed within the classification workflow. To support rule updates, we design an update engine with two complementary strategies that balance speed, accuracy, and robustness. The \textbf{Immediate Update} strategy directly inserts, deletes, or modifies rules within existing tuples, avoiding retraining and minimizing update latency. In contrast, the \textbf{Deferred Update} strategy incrementally trains or retrains the neural network, trading off update efficiency for higher model fidelity. Section~\ref{sec:update} provides a detailed description of these mechanisms.

\section{TaNG Classification and Updating}
\label{sec:op}

This section presents the TaNG classification workflow with post-verification mechanism and GPU acceleration, followed by two update strategies.

\subsection{Classification Workflow}
\label{sec:classification}

\subsubsection{Two-Stage Classification}
\label{sec:classi_process}

The classification process in TaNG is a two-stage procedure: model-based tuple prediction followed by intra-tuple rule search. 
The first stage is handled by the model. Each packet $P$ is represented as a vector of its field values $<p_1,p_2,...,p_F>$ and fed into the model, which outputs the predicted tuple index $tuple\_idx$ for the search engine.
The second stage is executed by the search engine: given $tuple\_idx$, it loads the corresponding tuple from the TSS structure. Using the tuple’s prefix signature, the engine truncates the relevant field value of $P$ and computes a hash value from the truncated values. Finally, the hash value is used to access the rules in the hash table, which are then compared one by one to return the action associated with the highest-priority matching rule.

TaNG’s classification model is inherently probabilistic and cannot guarantee classification accuracy. To mitigate this, we introduce a post-verification mechanism. When the model predicts an incorrect tuple, the search engine may encounter two scenarios. First, a matching rule with a lower priority is found within the predicted tuple. Second, no matching rule is found, a case that accounts for nearly all prediction errors in our experiments. Our post-verification mechanism primarily addresses the second scenario, whereas the first scenario will be addressed in future work (Section \ref{sec:limi}). Specifically, we set a flag to indicate whether a match rule is found in the predicted tuple. If none exists, an ordered search is performed across the remaining tuples until the highest-priority rule is located. This mechanism corrects the model’s prediction and significantly improves classification accuracy, as verified in Section \ref{sec:ex_infor}.

The classification time complexity of TaNG is $O(L \cdot N^2)$ dominated by floating-point computations. 
TaNG employs a fully connected network, which will be described in detail in Section \ref{sec:imple}.
Assuming the network has $L$ layers with $N$ neurons per layer, a single inference requires approximately $L \cdot N^2$ floating-point operations. Thus, the inference time complexity is $O(L \cdot N^2)$. Since it is independent of the number of rules, the inference performance remains stable. In contrast, the hash lookup within a tuple incurs a negligible overhead, with average complexity $O(1)$. In the worst case, all tuples may need to be searched, yielding a complexity of $O(t)$, where $t$ is the number of tuples. Since $t$ is typically much smaller than $N$, the overall classification cost is dominated by model inference, giving a total time complexity of $O(L \cdot N^2)$.

\subsubsection{GPU-accelerated Classification Framework}

The main performance bottleneck of learning-based methods such as TaNG lies in model inference, which is dominated by floating-point computations that CPUs handle inefficiently. By contrast, GPUs, with high floating-point throughput and natural parallelism, are inherently well-suited for such workloads. Motivated by these observations, we propose a CPU-GPU hybrid streaming classification framework for learning-based methods, where inference runs on the GPU, while post-inference search runs on the CPU. By streaming these stages across the two processors, the framework achieves efficient hardware–software co-design, thereby fully exploiting the potential of learning-based packet classification.

\begin{figure}
    \centering
    \includegraphics[width=1\linewidth]{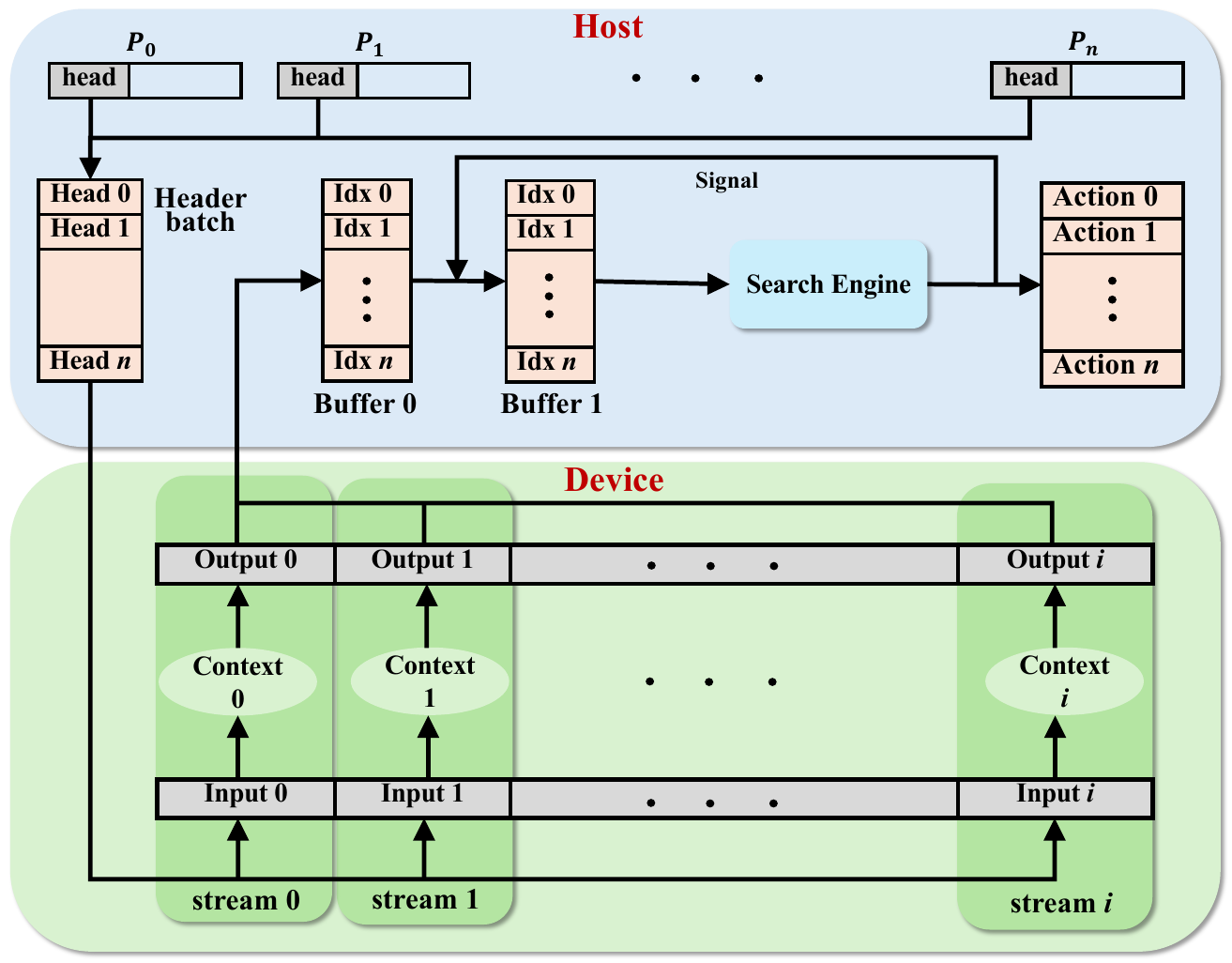}
    \caption{CPU-GPU hybrid classification framework.}
    \label{fig:pipeline}
\end{figure}

Figure~\ref{fig:pipeline} illustrates the CPU–GPU hybrid streaming classification framework, which splits responsibilities between the Host (CPU) and the Device (GPU). The Host manages data preparation and executes the search engine, while the Device is dedicated to model inference. Specifically, the Host extracts packet headers and organizes them into a contiguous memory block as a batch, which is then transferred to the GPU. On the Device, the batch is partitioned into $i$ segments, each assigned to a stream for asynchronous inference, with results stored in pre-allocated memory. As streams complete, their results are asynchronously copied back to the Host, where the search engine performs lookups and returns the final action.

Leveraging TensorRT \footnote{https://docs.nvidia.com/deeplearning/tensorrt/latest/index.html} and CUDA \footnote{https://docs.nvidia.com/cuda/index.html}, we simplify and accelerate GPU-side inference. TensorRT, NVIDIA’s high-performance inference engine, accelerates deep learning models through structural optimizations, precision quantization, operator fusion, and memory optimizations. Based on TensorRT, we can easily construct an optimized inference engine from the trained model, and subsequently instantiate multiple execution contexts deployed across different CUDA streams. Since CUDA streams inherently support asynchronous execution, this design enables flexible deployment while ensuring scalable throughput.

Since the CPU and GPU operate independently, we further overlap CPU lookups with GPU inference using a double-buffering scheme. As illustrated in Figure \ref{fig:pipeline}, the GPU writes results to $Buffer\ 0$ while the search engine reads from $Buffer\ 1$. During the search engine processing, buffer swapping is blocked, allowing the GPU to compute the next batch without delay. Once the search completes, the buffers are swapped, enabling continuous GPU inference and CPU search in a pipelined manner, maximizing resource utilization and overall throughput.


\subsection{Rule Update}
\label{sec:update}

TaNG provides two update strategies to balance update speed and classification performance: immediate updates and delayed updates. Immediate updates, based on the semi-structured model design in Section \ref{sec:semi-structure}, preserve the TSS structure and enable rapid rule update without modifying the model. Over time, however, accumulated updates may cause the actual tuple boundaries to drift from the model’s trained boundaries, inevitably degrading prediction accuracy. To address this, the delayed update strategy is introduced. While retraining the model is a straightforward solution, it incurs significant overhead. To balance efficiency and performance, delayed updates support both incremental training and full retraining, selecting the appropriate method based on search engine performance metrics and historical update information.

\subsubsection{Immediate Update}


In TaNG, immediate updates require the tuple set in the TSS structure to remain unchanged—no tuples can be added or removed. Under this constraint, modifying or deleting rules is straightforward: the rule can be directly edited or removed within its corresponding tuple. However, rule insertion requires a more careful strategy to ensure that each new rule maps to an existing tuple. To address this, we propose a restricted update strategy specifically for insertion scenarios.

Specifically, let $l^R = \{l_1^R, \ldots, l_F^R\}$ represents the prefix-length vector of a rule, where $l_i^R$ denotes the prefix length in the $i$-th field. Similarly, $l^T = \{l_1^T, \ldots, l_F^T\}$ represents the  prefix-length signature of a tuple. Rule insertion can then be divided into two cases. First, if a tuple exists such that $l_i^T = l_i^R$ for all $i$, the new rule is directly inserted into it. Otherwise, we search for tuples satisfying $l_i^T \le l_i^R$ for all $i$, and select the first tuple with the maximal $\sum_{i=1}^F l_i^T$ for insertion. For example, when inserting rule $R_{10}=\{X=100, Y=0**\}$ with prefix-length $(3,1)$ into the TSS structure built on Table \ref{tab:rules}, the candidate tuples are $T_3$, $T_4$, and $T_5$. Since $T_3$ has the maximal total prefix length and appears first, $R_{10}$ is inserted into $T_3$.


This insertion strategy means that the new rule's prefixes will be truncated to fit within the selected tuple, resulting in a suboptimal but valid mapping. 
However, this approach maximally leverages the rule's prefix information for hashing without altering the established TSS structure or adding new tuples. 
Overall, this immediate update strategy lays the groundwork for avoiding frequent model retraining, as it allows for model incremental training to handle certain shifts in the ruleset.


\subsubsection{Deferred Update}

Deferred updates in TaNG include two approaches—model retraining and incremental training—to more finely balance update cost and classification performance. Specifically, full retraining is straightforward: the TSS structure is rebuilt for the current ruleset, training data are generated (Section \ref{sec:train}), and the model is trained from scratch. Although this achieves optimal classification performance, it is time-consuming and often unnecessary when rules exhibit minor drift. To address this, incremental training offers a compromise by fine-tuning the existing model with new training data, enabling rapid adaptation. Notably, the immediate update strategy, which preserves the TSS structure, provides the foundation for incremental training: new data are generated on the existing TSS, and the model is fine-tuned accordingly. However, the timing of incremental training versus full retraining is critical, requiring a mechanism to control when and how updates occur.

In particular, ruleset drift directly affects classification performance; therefore, deferred updates are triggered by changes in classification throughput. We deploy a monitoring component on the search engine to continuously track throughput $th_{cur}$ over a predefined time window, while recording $th_{base}$, the throughput of the first window following the last full retraining. When $1 - th_{cur}/th_{base}$ exceeds a predefined threshold $\tau$, it indicates significant ruleset drift, triggering the deferred update.
The update engine then determines whether to perform incremental training or full retraining.
It first examines a counter tracking updated rules whose prefix length signature not match their assigned tuples since the last full retraining. If the proportion exceeds a threshold $\theta$, indicating many rules reside in suboptimal tuples and tuple boundaries may have significantly drifted, full retraining is triggered; otherwise, incremental training is performed. Through this mechanism, TaNG maintains high classification throughput while minimizing costly retraining, as validated in Section \ref{sec:dynamic_ex}.

\section{Neural Network Implementation}
\label{sec:imple}


This section details the implementation of the neural network model component in TaNG, covering the network design, training data generation, training procedure, and key hyperparameter settings.


\begin{figure}
    \centering
    \includegraphics[width=1\linewidth]{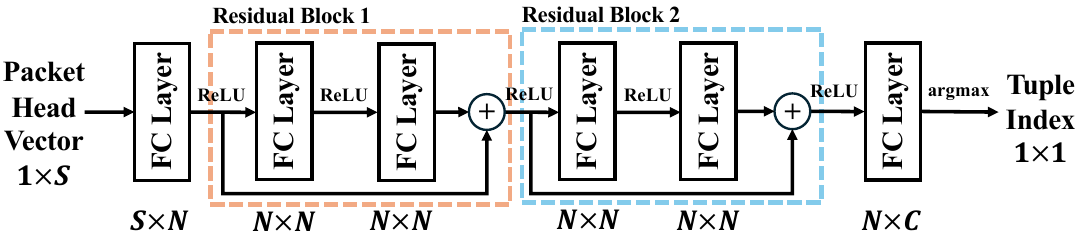}
    \caption{TaNG neural network architecture with 2 residual blocks, where $S$ denotes the input dimension, $N$ denotes the number of neurons per FC layer, and $C$ denotes the output dimension.}
    \label{fig:nn_architecture}
\end{figure}

\subsection{Neural Network Model}



The design of TaNG neural network model is guided by two key considerations. First, the model must possess sufficient capacity to fit tuple boundaries in high-dimensional space, thereby ensuring classification accuracy. Second, the model should not be overly complex; otherwise, excessive floating-point operations would increase inference cost and degrade classification throughput. Following these principles, we adopt a simple fully connected network and enhance its fitting capability with residual connections \cite{HeZRS16}. Specifically, since the constraint boundaries of tuples exhibit relatively regular structures in space, using a fully connected network to learn classification boundaries is feasible. However, despite the reduced overall complexity provided by TaNG’s semi-structured design, a plain multilayer perceptron (MLP) struggles with hundreds of classes. To address this, we introduce residual blocks, which deepen the network while mitigating overfitting. Furthermore, stacked residual blocks offer greater flexibility, making the model easier to adjust and extend.

Figure \ref{fig:nn_architecture} illustrates the TaNG network architecture with two residual blocks. The model begins with an initial fully connected (FC) layer to process inputs and ends with a final FC layer to generate outputs. The intermediate layers consist of residual blocks connected via skip connections, with each block containing two FC layers of $N$ neurons each and ReLU activations throughout. The input dimension $S$ is determined by the number of fields in the packet header (Section \ref{sec:train}), while the output dimension $C$ is determined by the TSS structure, equal to the total number of constructed tuples.



The operation of each residual block is defined as:
\begin{equation}
    B_i(x)=A(A(x\cdot w_1 + b_1)\cdot w_2)+b_2 + x)
\end{equation}
where $x$ is the input vector to the $i$-th block. The terms $w_1, w_2$ and $b_1, b_2$ are the weight matrices and bias vectors for the two FC layers within the block, respectively. The output of block $B_i$ serves as the input for the subsequent block $B_{i+1}$. The function $A()$ represents the ReLU activation, which is defined as:
\begin{equation}
    A(z)=\max(0,z)
\end{equation}
The final layer outputs class probabilities, from which the predicted tuple index is obtained via $argmax$.

\subsection{TaNG Training}
\label{sec:train}

TaNG training consists of two parts: data generation and model training.
To begin with, we define the model input format by segmenting packet headers into 16-bit chunks and converting each chunk into 32-bit floating-point values, which both preserves precision and facilitates neural network processing. For example, a classic 5-tuple header is split into 7 segments: the high and low 16 bits of SIP, the high and low 16 bits of DIP, the 16-bit SP, the 16-bit DP, and PRO. Based on this, we construct the training data format as $D = \langle s_1, s_2, \dots, s_k, tuple\_idx \rangle$, where $s_i$ is the $i$-th segment converted into a 32-bit floating-point value, $k$ is the number of segments, and $tuple\_idx$ is the tuple index of the rule matched in the TSS structure.

Building upon this format, data generation proceeds in two steps: raw data collection and training set preparation. We first process historical packets through the TSS lookup strategy, recording the tuple index $tuple\_idx$ of the highest-priority matching rule, thereby forming the raw dataset. However, since matched rules are often unevenly distributed across tuples, the raw dataset may exhibit significant class imbalance. 
To address this, we introduce a dynamically adjustable class-balance threshold $\alpha$. For any tuple class with fewer than $\alpha$ samples, we apply oversampling until its count reaches $\alpha$, thereby alleviating class imbalance. The value of $\alpha$ starts with an initial setting and then adjusted during training to ensure convergence.

Model training is implemented in PyTorch using the cross-entropy loss function and the Adam optimizer \cite{KingmaB14}. We set the batch size to 8192, train for 1000 epochs per round, and apply a learning rate schedule that decays the initial rate of 0.001 by a factor of 0.1 every 200 epochs. Throughout the training process, we continuously monitor accuracy while adapting the class-balance threshold $\alpha$ to ensure convergence. Specifically, after each 1000-epoch round, we evaluate the model’s accuracy $acc$ against a threshold $\beta$. If $acc < \beta$, we multiply $\alpha$ by 10 and regenerate the training set for the next round; otherwise, convergence is deemed successful and training terminates. Finally, the trained model is exported in ONNX \footnote{https://onnx.ai/onnx/intro/} format for deployment in the TensorRT inference engine.

\section{Evaluation}
\label{sec:eval}

This section first describes our experimental setup, followed by an evaluation of TaNG against two learned index-based methods and three traditional methods.

\subsection{Experiment Setup}
\label{sec:ex_set}

We compare TaNG against five representative algorithms: NuevoMatch(NM) \cite{NeuvoMatch2020}, NeuTree(NT) \cite{neutree2025}, PSTSS \cite{pfaff2015design}, CutSplit(CS) \cite{CutSplit2018}, and DBTable(DBT) \cite{dbtable2024}.
NM and NT serve as primary baselines representing state-of-the-art learning-based approaches.
The three traditional methods serve as baselines to highlight the performance ceiling of learning-based approaches on modern parallel hardware.
The rulesets are derived from the ClassBench-ng \cite{classbench-ng} suite, a widely recognized benchmark for packet classification algorithms, with 256k and 512k rules to assess scalability and performance on large-scale workloads.
We generate synthetic traffic that randomly matches the rules for each ruleset to measure raw throughput.
In addition, we use a real packet trace from WIDE\footnote{\url{https://mawi.wide.ad.jp/mawi/samplepoint-F/2024/202410141400.html}} to evaluate performance in real-world scenarios. Each traffic set comprises 1,000,000 packets.

NM is adapted from its open-source implementation, replacing the original training backend with PyTorch. The external classifier using TupleMerge \cite{TupleMerge2019} with paper-recommended settings.
We implement NT following the author's description, trained in PyTorch with bucket size 16, $EDF=64$, $RRF=0.3$, and $\gamma=0.4$. Since the models run on GPUs, the original binarized approximation computation method is not used. TaNG employs 6 residual blocks, with 512 neurons in each fully connected layer, setting $\alpha=1000$ and $\beta=0.95$. 

\begin{figure}[tb]
    \centering
    \subfigure[256k]{\includegraphics[width=1\linewidth]{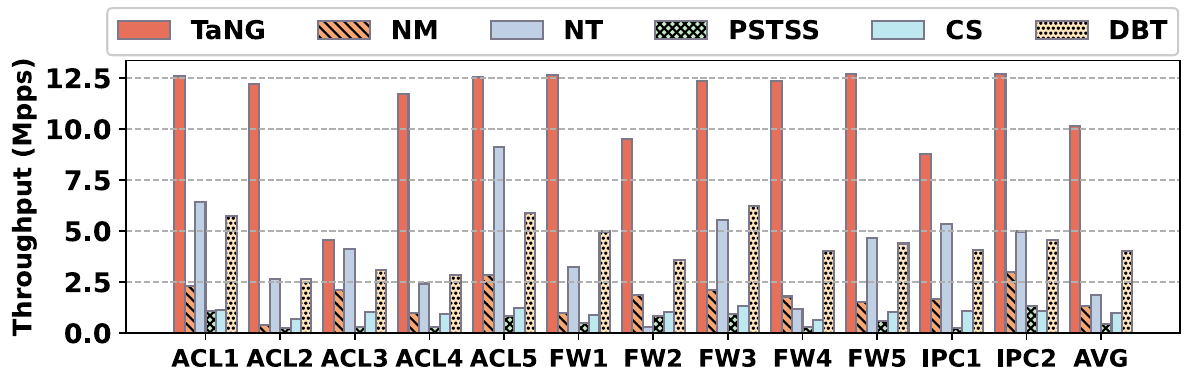}
    \label{lookup_256}}
    
    \subfigure[512k]{\includegraphics[width=1\linewidth]{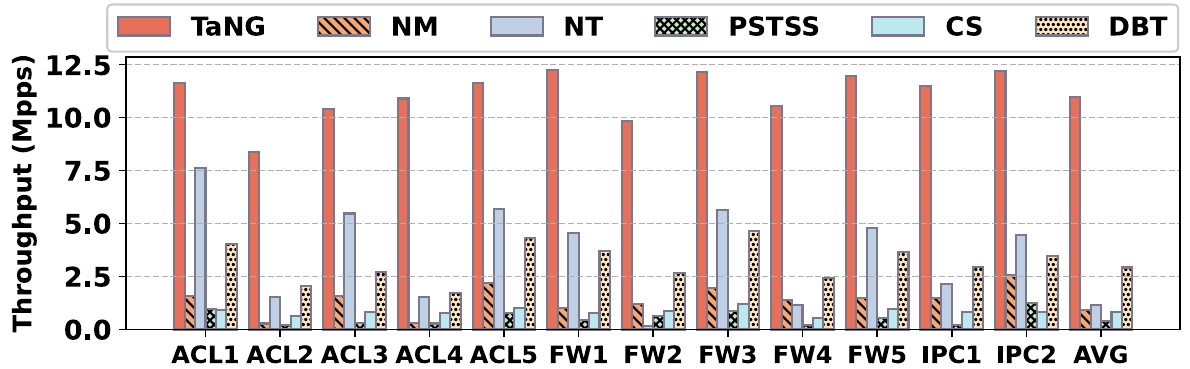}
    \label{lookup_512}}
    \caption{Throughput on synthetic traffic.}
    \label{fig:lookup}
\end{figure}

All experiments are conducted on a server with an AMD EPYC-7542 32-core CPU@2.9GHz, 252GB DRAM, an NVIDIA 3090 GPU, and Ubuntu 24.04. 
The models for NM, NT, and TaNG are trained using Python 3.12.7, PyTorch 2.5.0, and CUDA 12.8, and are then exported to ONNX for deployment. Classification is implemented in C++ using our CPU–GPU hybrid streaming framework. For PSTSS, CS, and DBT, we use their open-source implementations with paper-recommended parameters, all in single-thread mode. All classifiers are compiled with GCC 13.3 and -O3 optimization. Together, these baselines cover both learning-based and traditional approaches, providing a comprehensive evaluation of TaNG. All implementations are released as open source \footnote{https://anonymous.4open.science/r/TaNG}.

\subsection{Comparison}
In this section, we evaluate TaNG against five algorithms in terms of throughput, construction time, and memory consumption.

\begin{figure}[tb]
    \centering
    \subfigure[256k]{\includegraphics[width=1\linewidth]{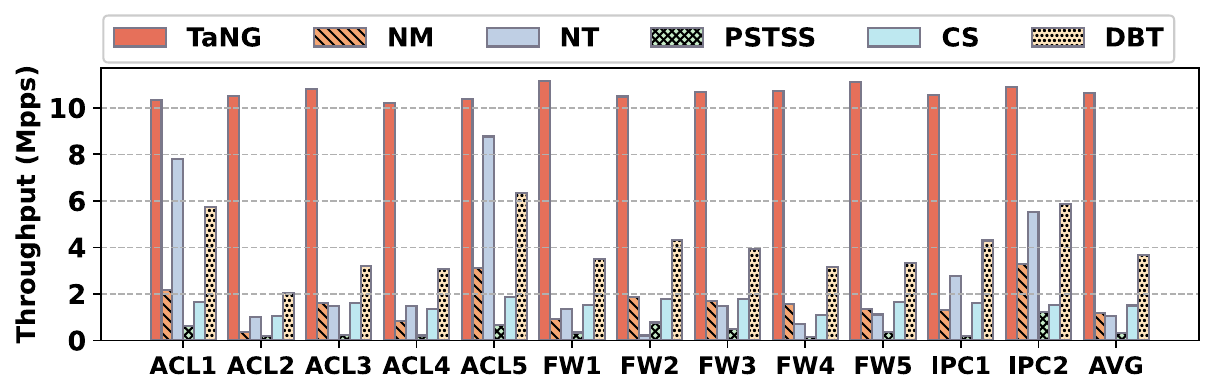}}
    \label{real_lookup_256}
    
    \subfigure[512k]{\includegraphics[width=1\linewidth]{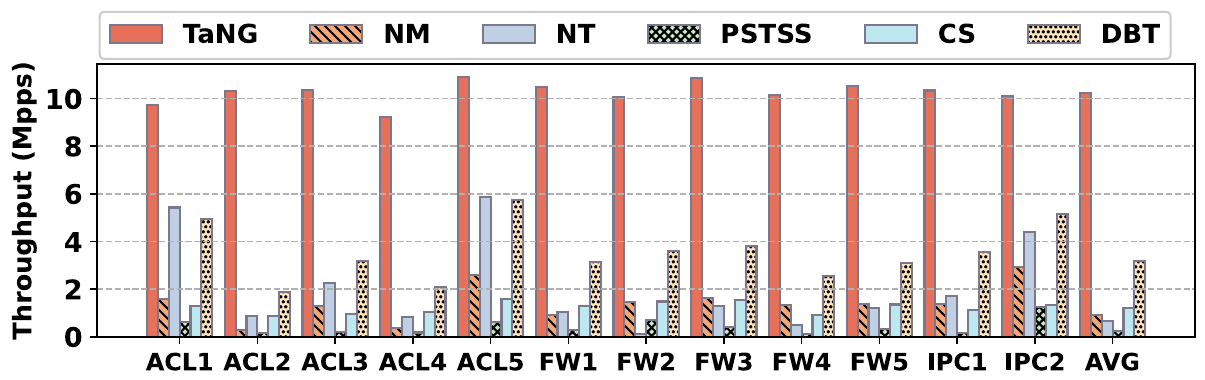}}
    \label{real_lookup_512}
    \caption{Throughput on real traffic.}
    \label{fig:lookup_real}
\end{figure}

\subsubsection{Throughput}


Throughput is a key metric for packet classification. We evaluate all algorithms under synthetic and real traffic. For TaNG, NM, and NT, we set the batch size to 8192 and use 4 CUDA streams for inference. 
Figure \ref{fig:lookup} shows the throughput on synthetic traffic, reported in millions of packets per second (Mpps).
TaNG achieves the highest throughput on both 256k and 512k rulesets, averaging 10.15Mpps and 10.98Mpps across 12 rulesets, corresponding to up to 7.64x and 12.19x speedup over NM, and 5.46x and 9.37x over NT. These gains mainly result from TaNG’s semi-structured model design, which converts classification into GPU-friendly floating-point operations while keeping CPU-side search efficient, as described in Section \ref{sec:ex_infor}. By contrast, NM and NT remain limited by CPU-side search and exhibit lower GPU acceleration due to the reminder set and rule replication, consistent with the analysis in Section \ref{sec:moti}.

Another advantage of TaNG is throughput stability. Since classification is almost entirely inference-bound, its throughput remains consistent across different rulesets and scales. In contrast, NM is limited by model coverage, and NT is affected by rule replication. For instance, NM achieves only 0.39Mpps and 0.28Mpps on acl2, while NT performs better on acl1 and acl5 where replication is low, consistent with the results shown in Figures \ref{fig:nm} and \ref{fig:nt}. Overall, TaNG attains up to 15.36x and 98.84x higher stability than NM, and 21.48x and 156.98x higher than NT on the 256k and 512k rulesets, measured in terms of the standard deviation of the average per-packet lookup time. Notably, the lower throughput of TaNG on the 256k ACL3 ruleset is due to relatively higher intra-tuple search overhead compared to other rulesets (\ref{sec:ex_infor}).


\begin{figure}[tb]
    \centering
    \subfigure[256k]{\includegraphics[width=1\linewidth]{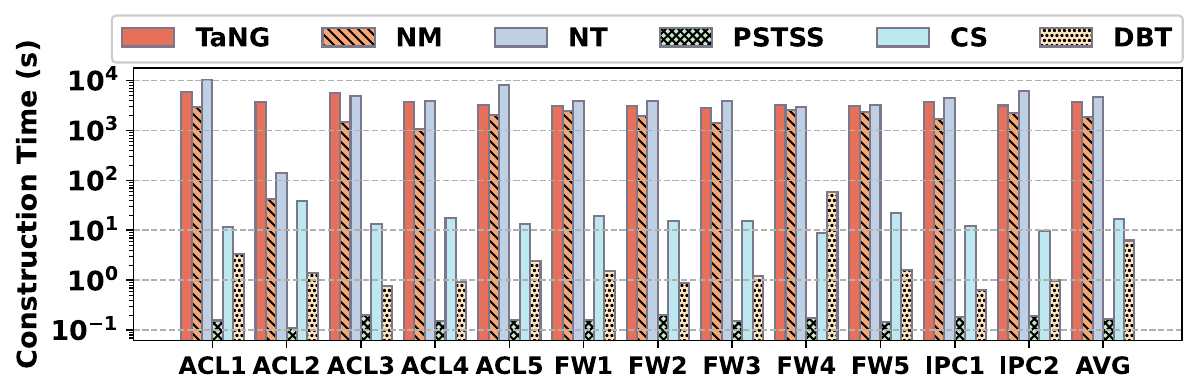}
    \label{construction_256}}
    
    \subfigure[512k]{\includegraphics[width=1\linewidth]{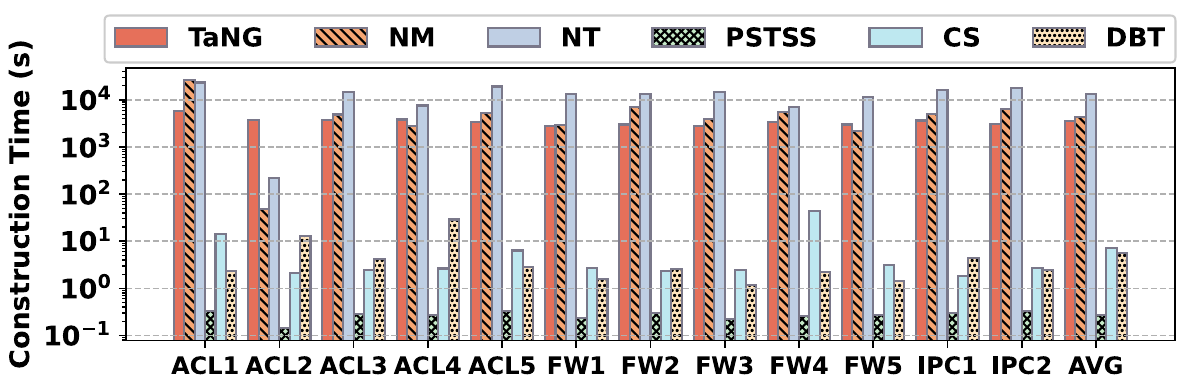}
    \label{construction_512}}
    \caption{Construction time.}
    \label{fig:construction}
\end{figure}


Compared to traditional methods, TaNG achieves significant gains through GPU-accelerated floating-point computation. On the 256k and 512k rulesets, it improves average throughput by 23.03x and 28.92x over PSTSS, 10.49x and 13.45x over CS, and 2.52x and 3.54x over DBT. PSTSS is limited by tuple traversal, which TaNG substantially reduces, while CS, as a decision-tree method, only visits a single leaf node and thus achieves higher throughput than PSTSS. DBT leverages feature extraction to map rules into a single hash table, significantly boosting throughput over PSTSS and surpassing CS. Although NM and NT exceed PSTSS and CS, their throughput lags behind DBT, highlighting both the potential and limitations of current learning-based methods. In terms of stability, TaNG also surpasses PSTSS, CS, and DBT by up to 146.29x, 22.8x, and 10.18x, respectively.

Figure \ref{fig:lookup_real} shows throughput under real traffic. TaNG maintains around 10Mpps across almost all rulesets, averaging 10.64Mpps and 10.23Mpps on 256k and 512k rulesets. At 256k, it achieves 8.95x, 10.31x, 34.98x, 7.09x, and 2.89x speedup over NM, NT, PSTSS, CS, and DBT; at 512k, the improvements rise to 11.28x, 15.14x, 39.88x, 8.6x, and 3.2x, demonstrating TaNG’s strong scalability for large-scale rulesets. TaNG also improves throughput stability by up to 259.92x, 472.66x, 718.81x, 50.45x, and 34.87x over NM, NT, PSTSS, CS, and DBT, confirming its consistently high throughput and stability under realistic workloads.


\subsubsection{Construction}
\label{sec:ex_constru}

Figure \ref{fig:construction} reports the construction time of all algorithms, including preprocessing. Overall, learning-based methods incur substantially higher costs than traditional ones. TaNG requires on average 61 minutes (256k) and 58 minutes (512k), compared to 30 and 73 minutes for NM, and 77 and 221 minutes for NT. Notably, TaNG achieves relatively stable costs across scales since its model parameter size is fixed and data generation is lightweight. By contrast, NM extracts disjoint rules subsets and NT performs complex bucket partitioning, making them sensitive to ruleset scale. Moreover, with least-squares fitting in place of neural network training, NT achieves extremely short training times of only 16.74s (256k) and 41.24s (512k).

Traditional methods, in contrast, exhibit construction costs that are orders of magnitude lower. PSTSS is the fastest, requiring only a single hash computation, with average times of 165 ms (256k) and 274 ms (512k). DBT follows, with 2.21s and 5.59s due to feature extraction and hashing, while CS requires 6.24s and 7.35s due to multiple rule partitioning steps. Unlike learning-based methods, traditional algorithms require no training and low preprocessing cost, enabling PSTSS, DBT, and CS to achieve near-instantaneous setup, while TaNG, NM, and NT still take tens of minutes overall. However, the construction overhead is typically justified by the substantial throughput gains.



\begin{figure}[tb]
    \centering
    \subfigure[256k]{\includegraphics[width=1\linewidth]{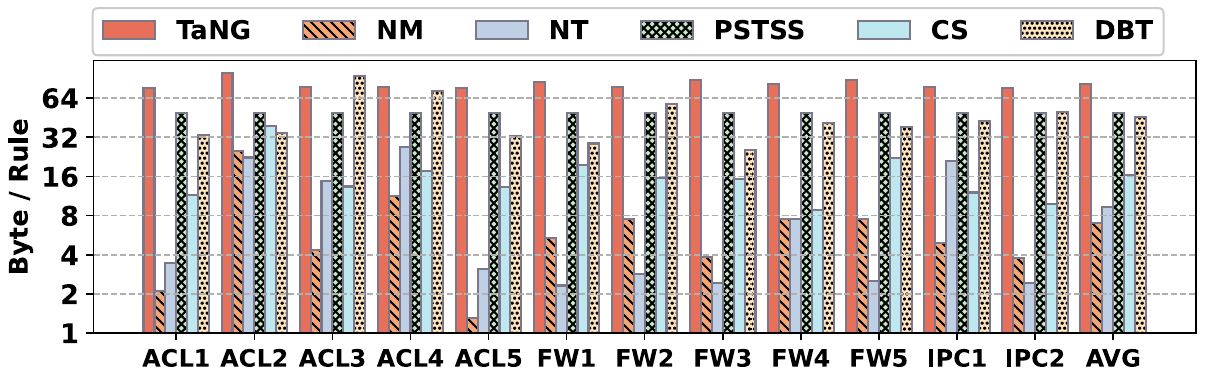}
    \label{mem_256}}
    
    \subfigure[512k]{\includegraphics[width=1\linewidth]{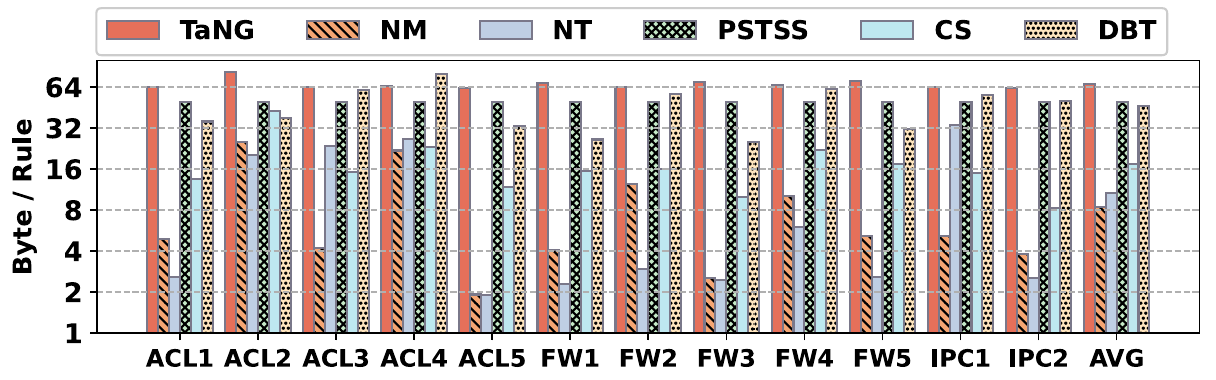}
    \label{mem_512}}
    \caption{Memory consumption.}
    \label{fig:Memory}
\end{figure}

\subsubsection{Memory}

Figure \ref{fig:Memory} reports memory usage in bytes per rule, including models, data structures, and rule pointers. NM and NT have the smallest footprints—7.06/9.31 B and 8.45/10.63 B per rule at 256k/512k—owing to lightweight models and structures, with NT slightly higher due to rule replication. Both are consistently lower than traditional methods, showing the compression advantage of them. Among traditional schemes, PSTSS and DBT each consume roughly 49 B per rule due to hash tables, while CS achieves lower usage (16.51/17.51 B) by reducing replication. TaNG consumes the most memory, as it preserves the full TSS structure plus an about 6.69 MB neural model, yielding 81.96 B and 66.82 B per rule at 256k/512k. Since the model cost of TaNG is almost constant, per-rule usage decreases with larger ruleset. Overall, although TaNG incurs the largest footprint, its throughput gains far outweigh this overhead.



\begin{figure}[tb]
    \centering
    \includegraphics[width=1\linewidth]{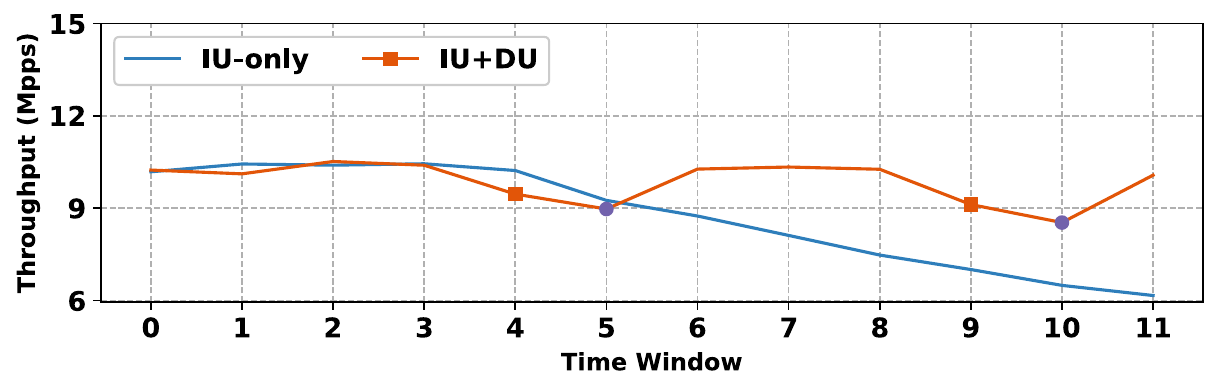}
    \caption{The throughput in the dynamic scenario with different update strategies.}
    \label{fig:shift_test}
\end{figure}

\subsection{Dynamic Experiments}
\label{sec:dynamic_ex}

To evaluate the effectiveness of TaNG’s update mechanism, we design a set of dynamic experiments. Starting with the 256k acl1 ruleset, in each update window we randomly delete 2,000 rules and insert 2,000 rules drawn from the other 11 rulesets. The update window size is set to 30 minutes. We evaluate two configurations: Immediate Update only (IU-only) and Immediate Update with Deferred Update (IU+DU). For IU+DU, we configure $\tau=5\%$, $\theta=10,000$, and cap incremental training time at 30 minutes.
A large $\tau$ hurts throughput stability, while $\theta$ is usually set high to avoid frequent retraining.
Figure~\ref{fig:shift_test} reports the average throughput of each time window under dynamic rule updates.


Under IU-only, TaNG achieves an average initial throughput of 10.18 Mpps and maintains stability for three windows. Throughput then declines at the fourth window and continues dropping, reaching 6.16 Mpps (a 39.48\% decrease). This shows that TaNG sustains stable performance under small updates even without model retraining.
By contrast, IU+DU mitigates throughput degradation. In our setup, retraining is not triggered. In this scenario, starting at 10.23Mpps, throughput drops to 9.45Mpps (7.62\%) at the fourth window, triggering deferred update.
After incremental training, the model is deployed in the fifth window, with throughput recovering to 10.26 Mpps by the sixth window and remaining stable for the next two windows. 
At the ninth window, throughput declines again to 9.11 Mpps, prompting another deferred update and subsequent recovery. Overall, the deferred update mechanism enables TaNG to adapt to ruleset drift, restoring throughput via incremental training whenever moderate changes occur.


\subsection{Statistical Information}
\label{sec:ex_infor}

To investigate the factors behind TaNG’s high and stable throughput, Table~\ref{tab:ana} and Table~\ref{tab:ana2} report four key statistics on the 256k and 512k rulesets with synthetic traffic: tuple prediction accuracy of the model, TaNG classification accuracy, the number of constructed tuples, and the average memory accesses per lookup in the search engine. The results show that the TaNG model achieves high tuple prediction accuracy—averaging 97.242\% on 256k rules and 96.529\% on 512k rules—ensuring that in most cases, only a single tuple needs to be accessed for classification. This significantly reduces the search engine overhead, offloading workloads from the CPU to the GPU, and explains why TaNG achieves substantially higher throughput compared to existing learning-based approaches.
Moreover, our post-verification mechanism significantly improves classification accuracy, demonstrating its effectiveness, with over 99.99\% achieved on both 256k and 512k rulesets. 

The number of tuples and the average memory accesses per lookup reflect the benefits of employing the TSS structure: it reduces the number of classes and enables efficient post-inference search. These properties are critical for lowering model complexity and ensuring that CPU-side search does not become a throughput bottleneck.
As shown in Table~\ref{tab:ana} and \ref{tab:ana2}, the average memory accesses per lookup are only 5.13 for the 256k ruleset and 4.14 for the 512k ruleset. Overall, these results demonstrate that TaNG concentrates most of the classification cost on GPU-accelerated model inference rather than the comparisons or logical operations on CPU. By combining a highly accurate neural network with a fast post-inference search engine, TaNG consistently delivers high throughput, even with large-scale rulesets.



\begin{table}[tbp]
\centering
  \caption{TaNG statistics on 256k rulesets.}
  \label{tab:ana}
  \resizebox{1\linewidth}{!}{
  \begin{tabular}{c| c c c c} \hline
  \textbf{Ruleset} & \textbf{Model Accu.} & \textbf{Classi. Accu.} & \textbf{Tuple Num.} & \textbf{Mem. Acce.} \\ \hline
  ACL1 & 93.117\% & 100\% & 179 & 2.75 \\
  ACL2 & 97.745\% & 99.999\% & 364 & 4.39 \\
  ACL3 & 94.468\% & 99.996\% & 320 & 15.06 \\
  ACL4 & 98.286\% & 99.998\% & 385 & 4.08 \\
  ACL5 & 99.876\% & 100\% & 138 & 2.32 \\
  FW1 & 99.615\% & 100\% & 144 & 3.48 \\
  FW2 & 90.491\% & 99.997\% & 69 & 6.51 \\
  FW3 & 99.708\% & 99.999\% & 97 & 2.77 \\
  FW4 & 98.535\% & 99.999\% & 87 & 6.01 \\
  FW5 & 99.393\% & 99.999\% & 124 & 3.31 \\
  IPC1 & 98.318\% & 100\% & 362 & 7.83 \\
  IPC2 & 97.349\% & 100\% & 31 & 2.99 \\ \hline
  \end{tabular}
  }
\end{table}


\section{Limitations and Future Work}
\label{sec:limi}

While TaNG delivers substantial throughput gains, it also faces several limitations. 
First, it cannot guarantee complete classification correctness. As discussed in Section \ref{sec:classi_process}, our post-verification mechanism only handles the second scenario. While classification accuracy exceeds 99.99\% in most cases, there remains a possibility of matching lower-priority rules, which limits its applicability in environments with strict correctness requirements. Second, although TaNG provides update mechanisms, it inherently trades off between update efficiency and classification performance. This constrains its use in scenarios demanding both real-time updates and consistently high performance.

Although in certain applications, such as network traffic monitoring and analysis \cite{wang2025sozenetworktelemetryneed}, matching a lower-priority rule is acceptable, our future work will still focus on two directions: ensuring classification correctness and optimizing update efficiency.
For the first challenge, a potential approach is to design a verification mechanism capable of quickly determining whether the matched rule is the intended one, and, if not, perform a global search to guarantee correctness. For the second challenge, we plan to explore faster model fine-tuning techniques to accelerate incremental training, enabling TaNG to adapt more rapidly to continuously evolving rulesets.



\begin{table}[tbp]
\centering
  \caption{TaNG statistics on 512k rulesets.}
  \label{tab:ana2}
  \resizebox{1\linewidth}{!}{
  \begin{tabular}{c| c c c c} \hline
  \textbf{Ruleset} & \textbf{Model Accu.} & \textbf{Classi. Accu.} & \textbf{Tuple Num.} & \textbf{Mem. Acce.} \\ \hline
  ACL1 & 89.294\% & 100\% & 183 & 2.69 \\
  ACL2 & 91.574\% & 99.994\% & 364 & 7.15 \\
  ACL3 & 96.94\% & 100\% & 322 & 4.65 \\
  ACL4 & 96.733\% & 99.999\% & 391 & 4.95 \\
  ACL5 & 96.458\% & 100\% & 139 & 3.65 \\
  FW1 & 99.799\% & 100\% & 144 & 3.45 \\
  FW2 & 93.589\% & 99.993\% & 71 & 5.2 \\
  FW3 & 99.745\% & 100\% & 97 & 2.57 \\
  FW4 & 96.281\% & 99.999\% & 87 & 6.59 \\
  FW5 & 98.929\% & 100\% & 124 & 3.65 \\
  IPC1 & 99.785\% & 100\% & 362 & 2.64 \\
  IPC2 & 99.217\% & 100\% & 31 & 2.46 \\ \hline
  \end{tabular}
  }
\end{table}

\section{Conclusion}
\label{sec:conclu}

This paper presents TaNG, a learning-based packet classification method optimized for high throughput. TaNG employs a semi-structured design that transforms existing learned models from a regression paradigm to a classification paradigm while integrating a traditional TSS data structure, effectively addressing limitations in model coverage and rule replication found in prior approaches. Leveraging our CPU–GPU hybrid streaming classification framework, TaNG achieves hardware–software co-design, significantly improving both classification throughput and stability. Experiments show that on 256k and 512k rulesets, TaNG achieves average throughput of 10.15Mpps and 10.98Mpps, corresponding to 7.64x and 12.19x speedup over NuevoMatch and 5.46x and 9.37x speedup over NeuTree. Moreover, we propose immediate and delayed update mechanisms for TaNG to balance update efficiency and classification performance. Our future work will focus on ensuring classification correctness and optimizing update responsiveness to further enhance TaNG’s applicability.




{\footnotesize \bibliographystyle{acm}
\bibliography{sample}}


\end{document}